\documentclass[aps,preprint,groupedaddress,longbibliography,nofootinbib]{revtex4-1}

\usepackage[utf8]{inputenc}
\usepackage{amsmath,amssymb}
\usepackage{amsfonts}
\usepackage{slashed}
\usepackage{xcolor}

\usepackage[pdftex]{hyperref}
\hypersetup{ 
pdfstartview=FitV,
colorlinks=true, 
linkcolor=black, 
citecolor=blue, 
filecolor=black, 
urlcolor=magenta,
}
\usepackage{graphicx}

\usepackage{tensor}
\usepackage{mathrsfs}
\usepackage{bbm}
\usepackage{graphicx}
\usepackage{subcaption}
\begin{document}
\title{A hydrodynamical description for magneto-transport in the strange metal phase of Bi-2201}
\author{Andrea Amoretti$^{\, 1,2}$, Martina Meinero$^{\, 1,3}$, Daniel K. Brattan$^{\, 2}$, Federico Caglieris$^{\, 4}$, Enrico Giannini$^{\, 5}$, Marco Affronte$^{\, 6}$, Christian Hess$^{\, 4,7}$, Bernd Buechner$^{\, 4,7}$, Nicodemo Magnoli$^{\, 1,2}$, Marina Putti$^{\, 1,3}$}
\affiliation{\textbf{1} Dipartimento di Fisica, Universit\`a di Genova,
via Dodecaneso 33, I-16146, Genova, Italy
}
\affiliation{\textbf{2} I.N.F.N. - Sezione di Genova, via Dodecaneso 33, I-16146, Genova, Italy}
\affiliation{\textbf{3} CNR-SPIN, Corso Perrone 24, 16152 Genova, Italy}
\affiliation{\textbf{4} Leibniz IFW Dresden, Helmholtz str. 20, D-01069 Dresden, Germany}
\affiliation{\textbf{5} Department  of  Quantum  Matter  Physics,  University  of  Geneva,24  Quai  Ernest  Ansermet,  CH-1211  Geneva,  Switzerland}
\affiliation{\textbf{6} CNR  Nano  Istituto  Nanoscience  -  sezione  S3 and Universit\'{a}  di  Modena  e  Reggio  Emilia - Dipartimento  di  Scienze  Fisiche,  Informatiche  e  Matematiche,  via  G.  campi  213/A,  41125  Modena, Italy}
\affiliation{\textbf{7} Faculty of Physics, Technische Universit\"{a}t Dresden, D-01062 Dresden, Germany}

\date{\today}
\begin{abstract}
High temperature superconductors are strongly coupled systems which present a complicated phase diagram with many coexisting phases. This makes it difficult to understand the mechanism which generates their singular transport properties. Hydrodynamics, which mostly relies on the symmetries of the system without referring to any specific microscopic mechanism, constitutes a promising framework to analyze these materials. In this paper we show that in the strange metal phase of the cuprates, a whole set of transport coefficients are described by a universal hydrodynamic framework once one accounts for the effects of quantum critical charge density waves. We corroborate our theoretical prediction by measuring the DC transport properties of Bi-2201 close to optimal doping, proving the validity of our approach. Our argument can be used as a consistency check to understand the universality class governing the behavior of high temperature cuprate superconductors.
\end{abstract}

\maketitle
Three decades after the discovery of high-$T_c$ superconductors \cite{bednorz1986} the underlying mechanisms responsible for their peculiar behavior remain mostly unexplained. The superconductive region of the phase diagram changes upon doping with electrons or holes into a Mott-insulating copper oxide; the chemical doping modifies the density of carriers of the CuO$_2$ planes, rapidly suppressing the anti-ferromagnetic order. Above the superconducting dome there exists a debated pseudo-gap phase whose origin is still enigmatic. For temperatures higher than the pseudo-gap temperature the appearance of a strange metallic phase, characterized by the unusual temperature dependence of the transport coefficients is one of the grand-unresolved-issues in these materials. A quantum critical point around the optimally doped region, whose presence would force the temperature dependence of the observables of the system to obey simple scaling laws, has been proposed to be the origin of the exotic phenomena measured in the strange metallic phase \cite{Sachdev1992}.

In recent years, a plethora of experiments 
\cite{traquada1995,fukita2002,abbamonte2005,Hoffman2002,wu2011,ghirardelli2012,chang2012,leboeuf2013,hucker2014,hashimoto2014,gerber2015,chang2016,cyr2018,ralevic,neto2014,loretl,keimer2015,Arpaia2019}, using a variety of distinct experimental techniques, have proven that charge density wave (CDW) order is an ubiquitous feature of the phase diagram of all cuprate superconductors. In particular, resonant X-ray scattering measurements on Bi-2201 \cite{peng2018} have shown that two-dimensional CDW order is not only present in the under-doped regime, but it persists through optimal doping into the over-doped region of the phase diagram, suggesting that CDW order might play a dominant role in determining the transport properties of these materials as well.

From the theoretical point of view, the coexistence of many intertwined phases makes it difficult to develop reliable microscopic models to explain the behavior of high temperature superconductors (see e.g. \cite{RevModPhys.87.457} for a review). Recently, studies that try to explain the electric transport properties of these materials relying only on the hydrodynamic universality class, without referring to any specific microscopic mechanism, have appeared \cite{Delacretaz:2016ivq,Delacretaz:2017zxd,Delacretaz:2019wzh}. 

Hydrodynamics is a tool which gives an accurate description of any interacting system, classical or quantum, as long as the fundamental interacting degrees of freedom reach local thermal equilibrium fast. To observe hydrodynamic effects from electrons in a solid, the interaction time must be the fastest time scale in the system. In normal Fermi liquid systems however, the quasi-particles interact with each other very weakly, making the observation of the hydrodynamics of electron fluids notoriously hard. The situation changes in systems such as high temperature superconductors, where the strong interactions between electrons make the quasi-particle Fermi liquid picture unreliable. In these materials the only long-lived excitations are associated with (almost-)conserved quantities, like the charge current and the energy current, making hydrodynamics the ideal framework to describe the relaxation of these modes. Then, contrary to the Boltzmann equation approach, hydrodynamics does not require the existence of quasi-particles. 

The hydrodynamic approach has been used successfully to analyze the properties of graphene \cite{sachlu} and also very recently in new strongly coupled materials \cite{johan}. In \cite{Delacretaz:2016ivq,Delacretaz:2017zxd,Delacretaz:2019wzh}, using a hydrodynamic approach, fluctuating CDW order was proposed as an explanation for the off-axis peak in the optical conductivity developing at high temperature in many cuprates, including in particular BSCO and BSCCO \cite{Tsvetkov1997,Hwang2007}. Moreover, there is experimental evidence indicating that not only electric but also the thermo-electric transport, for example the Nernst coefficient \cite{taillefer2009}, is highly affected by the presence of the CDW order, not only at the fluctuating level but also closer to the superconducting phase transition. 

Based on this idea, in this paper we included in the hydrodynamic theory of a two-dimensional CDW the effects of an external magnetic field $B$ perpendicular to the CDW $\{x,y\}$-plane. We have analyzed how the magnetohydrodynamic approach constrains the entire set of electric, thermo-electic and thermal DC transport coefficients. In this approach, parity is broken solely by the presence of $B$, which can be seen as an external potential. We refer to the Supplemental Material (SM) for technical details. As already found in \cite{Delacretaz:2016ivq}, the electric conductivity at low $B$ splits into two terms, an incoherent conductivity $\sigma_0$ which characterizes the electric transport at the quantum critical point, and an additional term $\tilde{\sigma}$ which includes the relaxation mechanisms, 
\begin{equation}\label{conduDC}
\sigma_{\mathrm{DC}}=\sigma_0+\tilde{\sigma}\qquad \text{with} \qquad \tilde{\sigma}=\frac{n^2}{\chi_{\pi \pi}}\frac{\Omega_1}{\Omega_1 \Gamma+\omega_0^2} \ ,
\end{equation}
where $\Gamma$ is an external momentum relaxation rate which takes into account any external mechanism of momentum loss by the electronic plasma (e.g.~the interactions with the underlying lattice), $n^2/\chi_{\pi \pi}$ is the Drude weight and $\omega_0$ is the so called pinning frequency \cite{PhysRevB.17.535,PhysRevB.18.6245,PhysRevB.19.3970}. The pinning frequency is a characteristic frequency for the CDW collective modes (pseudo-Goldstone bosons) and it is non-zero as soon as the they acquire a small mass e.g.~in the presence of impurities or disorder. The additional parameter $\Omega_1$ is a phase relaxation for the collective modes associated to the CDW. Initially $\Omega_{1}$ was included to account for the consequences of topological defects \cite{gruner88} but, as recently noted in \cite{Amoretti:2018tzw,Amoretti:2019cef}, it also gives a non-trivial contribution to the transport when the CDW pseudo-Goldstone bosons have a small mass, namely in the presence of simple disorder (similarly to $\omega_0$). For $\Omega_1=\sigma_0=0$, one finds that $\sigma_{\mathrm{DC}}=0$, as already known from the standard CDW approach \cite{PhysRevB.17.535,PhysRevB.18.6245,PhysRevB.19.3970}. However, in the most general case with $\Omega_1,\sigma_0\neq0$, hydrodynamics allows for a CDW with metallic behavior. Importantly, and in contrast with the standard Drude scenario, as long as $\Omega_1, \, \omega_0 \neq 0$, the DC conductivity is finite even if $\Gamma=0$. 

The other DC transport coefficients are tightly constrained in terms of a small number of parameters. Specifically, the low-$B$ behavior of the DC electric resistivity $\rho_{xx}$, the Hall angle $\cot \Theta_H=\sigma_{xx}/\sigma_{xy}$, the magnetoresistance  $\Delta \rho/\rho$, the Nernst coefficient $N$ and the thermal Hall conductivity $\kappa_{xy}$ are of the following form:
\begin{subequations}\label{tranco}
\begin{align}
&\rho_{xx}=\frac{1}{\sigma_0+\tilde{\sigma}}+\mathcal{O}(B^2), \label{rhoxxeq}\\
&\frac{\Delta \rho}{\rho}=B^2 \frac{\sigma_0^3 \ \tilde{\sigma}}{n^2}\frac{1}{(\sigma_0 +\tilde{\sigma})^2}+\mathcal{O}(B^4), \label{magreeq}\\
&\kappa_{xy}=-B T \frac{\tilde{\sigma}^2  s}{n^4}\left(n s+2 \frac{\alpha_0 n^2}{\tilde{\sigma}}\right) +\mathcal{O}(B^3), \label{themralhalleq}\\
&\cot \Theta_H =\frac{n}{B \tilde{\sigma}}\frac{1+\frac{\sigma_0}{\tilde{\sigma}}}{1+2 \frac{\sigma_0 }{\tilde{\sigma}}}+\mathcal{O}(B), \label{hallangleeq} \\
&N=\frac{B \ \sigma_0 \  \tilde{\sigma}}{n^2(\sigma_0+\tilde{\sigma})^2}(s \sigma_0-n \alpha_0)+\mathcal{O}(B^3), \label{nernsteq}
\end{align}
\end{subequations}
where $\sigma_0$ and $\tilde{\sigma}$ are the incoherent and relaxation conductivities defined in \eqref{conduDC}, $T$ is the temperature and $s$ and $n$ are the entropy density and the charge carrier density respectively. The parameter $\alpha_0$ is an incoherent thermo-electric conductivity. It can be expressed as a function of the chemical potential $\mu$ and $\sigma_0$, namely $\alpha_0 \sim - \mu \sigma_0/T$, if we assume that the low $T$ behavior of the system is influenced by the presence of a relativistic covariant quantum critical point (see the SM). This implies that the theory of quantum critical CDW hydrodynamics allows us to express the five transport coefficients in \eqref{tranco} as a function of only four temperature dependent quantities: $\sigma_0$, $\tilde{\sigma}$, $n$ and $s$. Eventually, by measuring four of the quantities in \eqref{tranco} one can predict the temperature dependence of the fifth transport coefficient.

This puts hydrodynamics on a privileged ground to analyze strange metals. In fact, being agnostic on the specific microscopic mechanism responsible for the peculiar behavior of these materials, hydrodynamics provides a way to test the appropriate universality class to which these systems belong at the level of the DC transport properties. 

To test the last statement, we have measured the five DC magneto-transport coefficients \eqref{tranco} in Bi$_2$Sr$_2$CuO$_6$. Being electrical and/or transverse transport coefficients, these quantities are largely independent of the effects of lattice phonons and constitute ideal observables to test the electronic properties of the material. Moreover, measuring all the transport coefficients in the same sample allows us to be confident that all the data refer to the same doping and impurity level. By fixing the temperature dependence of the four phenomenological quantities in \eqref{tranco} using the low temperature behavior of $\rho_{xx}$, $\Delta\rho/\rho$, $\cot \Theta_H$ and $\kappa_{xy}$, we have been able to uniquely determine the low temperature dependence of the Nernst coefficient. The obtained result is in strong agreement with our experimental observations.

\begin{figure*}
	\centering	
	\includegraphics[width=1\linewidth]{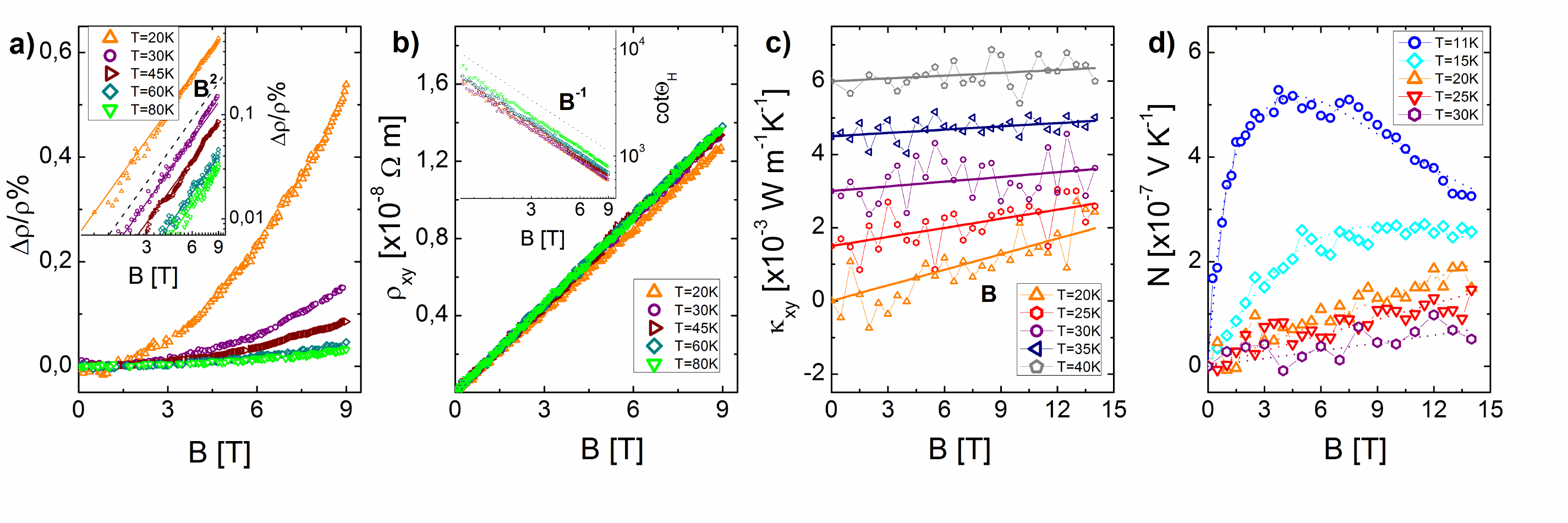}
	\caption{Magnetic field dependence of $\Delta\rho/\rho$ {\bf a$)$}, $\rho_{xy}$ {\bf b$)$}, $\cot\Theta_H$ (inset of figure {\bf b$)$}), $\kappa_{xy}$ together with its linear fits (solid lines) {\bf d$)$} and $N$ (dashed lines are guides to the eye) {\bf c$)$}. In {\bf d$)$} the curves are shifted to avoid overlapping.}
	\label{fig1}
\end{figure*}

\section{The experimental results}
Among the cuprates, Bi$_2$Sr$_2$CuO$_6$ (Bi-2201) is one of the most promising materials to test the magneto-hydrodynamic scenario, due to its relatively low critical temperature ($T_c\sim 10-20$ K) \cite{fleming1991,khasanova1995,piriou2010} which allows for a systematic study of its normal state properties down to low temperatures \footnote{It has been reported that similar compound might have complicated Fermi surface, presenting pockets \cite{king2011}, even though the relevance of these pockets on the transport properties remains unclear.}. In this compound, the main signatures of quantum criticality have been reported: its electric resistivity is $T$-linear up to 700 K \cite{martin1990} and CDW modulations have been observed in a wide region of its phase diagram \cite{peng2018}.  

We performed our measurements on single crystals of Bi-2201 close to optimal doping. The electric characterization has been undertaken with several samples and we report here the data relative to three crystals, named L2, K8 and K10, whereas the thermal and thermoelectric characterizations have been performed on the L2 sample (its size being suitable for these measurements). 
The in-plane resistivity $\rho_{xx}$ of the L2, K8 and K10 samples as a function of temperature shows a quite sharp superconducting transition at $T_c\sim 11$~K and, above this temperature, $\rho_{xx}(T)$ is almost linear up to 380~K (see the SM for details).

Figure \ref{fig1}a displays the magnetoresistance, defined as $\Delta\rho/\rho=(\rho_{xx}(B)-\rho_{xx}(0))/\rho_{xx}(0)$, of the L2 sample as a function of $B$ up to 9 T applied in the out-of-plane direction, for selected temperatures in the range 20-80 K. As evidenced by the bi-logarithmic scale reported in the inset, $\Delta\rho/\rho$ increases as $B^2$, as expected for the orbital contribution. The values are quite small (lower than 0.5\%) and they abruptly decrease with temperature.

In figure \ref{fig1}b we report the transverse resistivity $\rho_{xy}$ of the L2 sample as a function of $B$ up to 9 T applied in the out-of-plane direction, at selected temperatures in the range from 20 to 80 K. $\rho_{xy}$ is $B$-linear and it exhibits a weak temperature dependence. The inset of figure \ref{fig1}b shows the bi-logarithmic plot of the inverse Hall angle $\cot\Theta_H=\rho_{xx}/\rho_{xy}$ which is proportional to $B^{-1}$ as a consequence of the small magnetoresistance and the linear dependence of $\rho_{xy}$. The same behavior has been measured also for K7 and K10 samples. 

Figure \ref{fig1}c shows the magnetic field dependence up to 14 T of the thermal Hall conductivity $\kappa_{xy}$ measured at selected temperatures in the range 20-40 K (see the SM). It grows linearly with $B$, reaching values lower than $2\times10^{-3}$~W m$^{-1}$K$^{-1}$ measured at 20~K and 14 T (solid lines are linear fits to the data and the curves are shifted to avoid overlapping) with a pretty small slope, which monotonically decreases with increasing temperature. Above 40 K, $\kappa_{xy}$ is vanishingly small, so that no meaningful signal can be extracted from the noise.

In figure \ref{fig1}d we report the magnetic field dependence of $N$ from 0 to 14 T at $T=11, 15, 20, 25, 30$ K. At 11 K, just above $T_c$, $N$ shows the characteristic “tilted-hill” profile which has been widely ascribed to contributions from superconducting vortices \cite{wang2006}. At $T=15$ K, the vortex contribution is still visible, even if markedly reduced. For $T\geq20$ K, $N$ is $B$-linear with a slope which slowly decreases with increasing $T$. We therefore exclude any vortex contribution above 20 K where we still observe a non-negligible Nernst signal, already reported in cuprates, whose origin remains controversial \cite{wang2006,Hess2010}. 

The magnetohydrodynamic expansion for a parity invariant system gives that the transport coefficients behave as reported in equations \eqref{tranco}, namely $\Delta\rho\sim B^2$, $\cot\Theta_H\sim B^{-1}$, $\kappa_{xy}\sim B$ and $N\sim B$ (see SM). As shown in figure \ref{fig1}, these predictions agree with the experimental results above 20 K, since below this temperature the Nernst effect is influenced by vortex contributions.

\begin{figure*}[t]
	\centering	
	\includegraphics[width=1\linewidth]{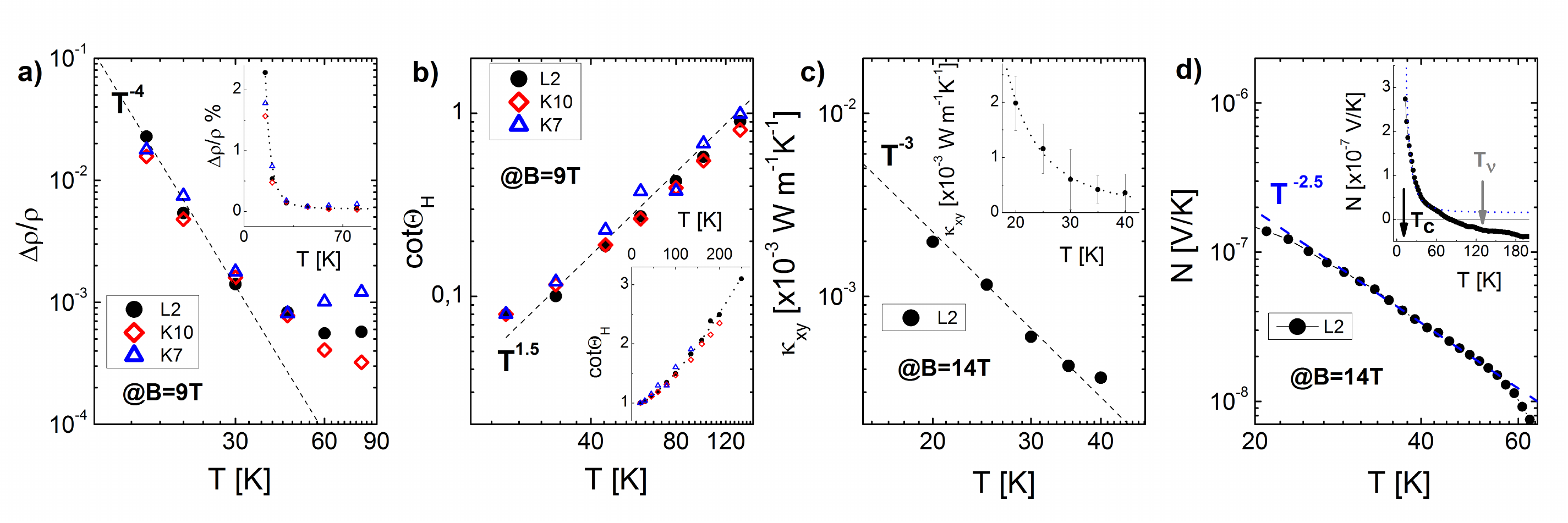}
	\caption{Bi-logarithmic plot of $\Delta\rho/\rho$ {\bf a$)$}, $\cot\Theta_H$ {\bf b$)$}, $\kappa_{xy}$ {\bf c$)$} and $N$ {\bf d$)$} vs $T$, together with functions which well reproduce the data (dashed lines) for L2, K7 and K10 samples. Linear plots of the data are reported in the insets.}
	\label{fig2}
\end{figure*}

In addition to the magnetic field scalings, we have analysed the temperature dependence of $\Delta\rho/\rho$,  $\cot\Theta_H$, $k_{xy}$ and $N$. Figure \ref{fig2}a displays the bi-logarithmic plot of $\Delta\rho/\rho$ at $B=9$ T for selected temperatures in the range 10-100 K. We note that the data relative to the L2, K10 and K7 samples almost overlap between 15 and 40 K, where $\Delta\rho/\rho$ decreases by about two orders of magnitude, as fast as $\sim T^{-4}$. Above 40 K, $\Delta\rho/\rho$ is extremely small and the data relative to the three samples are much scattered so that it is hard to draw any conclusion about the temperature dependence for $T>$40 K.

Figure \ref{fig2}b shows the bi-logarithmic plot of $\cot\Theta_H=\rho_{xx}/\rho_{xy}$ at $B=9$ T for selected temperatures in the range 15-150 K of the L2, K10 and K7 samples. This plot provides evidence for the good reproducibility of the data, which fall on a line $T^{1.5}$.  
 
Figure \ref{fig2}c shows the bi-logarithmic plot of $k_{xy}$ between 20 K and 40 K at $B=14$ T. It decreases by rising $T$ as fast as $T^{-3}$. 
 
The inset of figure \ref{fig2}d shows the temperature dependence of the Nernst coefficient $N$ at $B=14$ T, in the range 15-200 K.  $N$ is small and negative at high temperatures, whereas it starts to be enhanced towards positive values for $T<T_\nu \sim 130$ K, $T_\nu$ (marked with a grey arrow in inset of figure \ref{fig2}d) being the onset of the upturn in $N$ (see Methods for the estimation of $T_\nu$). 

Figure \ref{fig2}d shows the bi-logarithmic plot of $N$ in the temperature range 20-60 K (black dots) together with the function $T^{-2.5}$ which reproduces the temperature dependence of the data above 20 K and below 60 K. Indeed, as discussed above, below 20 K the vortex contribution prevails, whereas above 60 K, $N$ changes sign (see the inset), indicating that another mechanism comes into play.

In summary, in the temperature range 20 K$<T<$60 K where the system is far enough from the superconducting transition and the thermal and thermoelectric observables are significantly different from zero, the temperature behaviors for the measured properties are the following: $\rho_{xx}\sim T$, $\Delta\rho/\rho\sim T^{-4}$, $\cot\Theta_H\sim T^{1.5}$, $\kappa_{xy}\sim T^{-3}$ \footnote{As already mentioned, it was not possible to separate $\kappa_{xy}$ from the noise above 40 K due to the vanishingly small value of the signal. However we tentatively extend up to 60 K the range of validity of $\kappa_{xy}\sim T^{-3}$ to test if a consistent scenario including all the transport coefficients can be drawn.} and $N\sim T^{-2.5}$.   

\section{Matching experiment with hydrodynamic predictions}

The only consistent way to fit the behavior of the resistivity $\rho_{xx}$ \eqref{rhoxxeq} and the Hall angle $\cot \Theta_H$ \eqref{hallangleeq} with the hydrodynamic prediction is that the quantum critical incoherent conductivity $\sigma_0$ is dominating $\rho_{xx}$ at low $T$. On the other hand, $\cot \Theta_H$ is uniquely  determined by the ratio $n/\tilde{\sigma}$ which can be fixed separately. Eventually, $\rho_{xx} \sim \frac{1}{\sigma_0} \sim T$, $\cot \Theta_H \sim \frac{n}{B \tilde{\sigma}} \sim T^{1.5}$. This is a known hydrodynamic mechanism \cite{Blake:2014yla}. In contrast to the Fermi liquid, the fact that the electric conductivity splits into two terms ($\sigma_{DC}=\sigma_0+\tilde{\sigma}$), allows for two different mechanisms to determine the behavior of the resistivity and the Hall angle. In this case, the incoherent quantum critical conductivity is dominating the electric longitudinal transport while the relaxation conductivity $\tilde{\sigma}$ is determining the temperature dependence of the Hall angle.

Imposing the temperature behavior of the magnetoresistance in \eqref{magreeq}, $\Delta \rho/\rho \sim T^{-4}$, fixes the scalings of $n \sim T^{1.5}$ and $\tilde{\sigma} \sim T^{0}$ separately. The charge carrier density $n$ is decreasing with $T$, making the incoherent part of the conductivity $\sigma_0$ dominant at low temperatures.

To fix the temperature dependence of the entropy density $s$ we use $\kappa_{xy} \sim T^{-3}$ \eqref{themralhalleq}. In particular:
\begin{equation}\label{scalingk}
\kappa_{xy} \sim \mu B \frac{\sigma_0 \ \tilde{\sigma}}{n^2} s \sim T^{-3} \qquad \Rightarrow \qquad s \sim T \ ,
\end{equation}
where we have imposed that $\alpha_0 = -\mu \sigma_0/T$ close to the quantum critical point. In the SM we report specific heat measurements of the BSCO crystal, which confirm the linear dependence with temperature of the electronic specific heat, in accordance with \eqref{scalingk}. Remarkably, an entropy linear in temperature is in accordance with the early measurements of the specific heat on optimally doped YBCO performed by Loram et al. \cite{Loram}.

Having fixed all the four quantities appearing in \eqref{tranco}, the temperature dependence of the Nernst coefficient is now fully determined: Specifically:
\begin{equation}\label{scalingN}
N \sim \frac{\mu B \tilde{\sigma}}{n T} \sim \frac{\mu}{T \cot \Theta_H} \sim T^{-2.5} \ .
\end{equation}
Remarkably, equation \eqref{scalingN} perfectly agrees with the experimentally observed temperature dependence of $N$ between 20 and 60 K (figure \ref{fig2}d). Analogously to the Hall angle, $\tilde{\sigma}$ completely determines the behavior of $N$. It is important to note that in ref.~\cite{taillefer2009} the enhancement of $N$ below $T_\nu$ has been interpreted as a signal of Fermi surface reconstruction due to the onset of charge order, which is measured to set in at $\sim T_\nu/2$. For our system this would imply that CDW order is the dominant mechanism for $T<65$ K,  suggesting that the CDW parameters $\Omega_1$ and $\omega_0$ should dominate over $\Gamma$, so that $\tilde{\sigma}=\frac{n^2}{\chi_{\pi \pi}}\frac{\Omega_1}{\omega_0^2}$. In order to fulfill the scaling previously determined $\Omega_1/\omega_0^2 \sim T^{-3}$, which can be achieved if $\Omega_1\sim T$, $\omega_0\sim T^2$. In a scale invariant theory governed only by a relativistic quantum critical point one would expect $\Omega_1\sim\omega_0\sim T$; here however $\omega_0$ is proportional to the mass of the CDW pseudo-Goldstone boson (see SM) which defines an additional scale allowing for $\Omega_1\sim T$, $\omega_0\sim T^2$. In the opposite scenario with $\Gamma$ being the dominant scale, one find $\tilde{\sigma}=\frac{n^2}{\chi_{\pi \pi}\Gamma}$, implying $\Gamma\sim T^3$. This would be compatible in a clean and multi-band metal with electron-phonon scattering being the dominant mechanism, which is not our case.

\section{Outlook}
This paper provides a consistency check, based on the analysis of DC magneto-transport coefficients, of hydrodynamics being the correct framework to describe the transport properties of the strange metal phase of cuprates. Hydrodynamics constrains the whole set of transport coefficients in a tight way, and implies that both the thermodynamic properties and the relaxation rates of the electronic system must have a very specific behavior in order to fit the model. Two main future directions are at hand. 
Firstly, even though evidence has been reported to suggest that $\tilde{\sigma}$ is dominated by the CDW, comparing optical measurements of the electric conductivity against hydrodynamic predictions will allow us to make a conclusive statement.
 
Secondly, it is worth mentioning that for other systems/doping levels, different scaling behaviors for the transport coefficients are reported. $\cot \Theta_H$ has been observed to scale as $T^2$ in other cuprates \cite{ando1999,coleman1996}, and different temperature dependencies of $\kappa_{xy}$ have been reported for overdoped Nd-/Eu-LSCO \cite{grissonnanche2019} and optimally doped YBCO \cite{zhang2000,matusiak2009}. Also, the electronic specific heat of YBCO has been observed to be doping dependent \cite{Loram}. The measurements of the complete set of transport properties in these and other materials, performed again on the same sample, will allow us to better test the range of validity of the hydrodynamic approach and to make a firm statement on the mechanism governing the transport properties of cuprates in different regions of the phase diagram.

\section{Acknowledgments}
We acknowledge discussions with I. Pallecchi and G. Lamura. AA would like to thank Daniel Arean, Blaise Gouteraux, Daniele Musso and Paolo Solinas for countless discussions on related topics. The work of MM has been partially supported by the DAAD Scholarship, Research Grants Award – Short-term grants, 2017 (57314023). 

\appendix
\section{Theoretical framework: 2D charge density wave magneto-hydrodynamics}
\label{theometh}
In this section we will derive the hydrodynamic theory of the two dimensional CDW including the effects of an external magnetic field perpendicular to the CDW plane $\{x,y\}$, following the analysis performed in \cite{Delacretaz:2016ivq,Delacretaz:2017zxd,Delacretaz:2019wzh}. 

Our model will consist of the typical ingredients for describing the hydrodynamics of a charged fluid (see e.g.~\cite{Kovtun:2012rj}); with the addition of two pseudo-Goldstone bosons, describing the spontaneous breaking of translations in the $x$ and $y$ directions. The hydrodynamic expansion requires one to systematically add derivatives of the thermodynamic sources, respecting any symmetries, to the ``constitutive relations'' - which are expressions for the currents of each conserved charge. We shall work to order one in derivatives and linearise about a constant, global background temperature $T$ and chemical potential $\mu$. This is sufficient to define the transport coefficients discussed in the main part of the present paper.

 To start this procedure requires us to supply the form of the currents and the associated conservation equations in global thermodynamic equilibrium. In particular we need to understand how the pseudo-Goldstone modes contribute to these expressions. The free energy ($F$) of our pseudo-Goldstone bosons $(\phi_{i=1,2})$ to quadratic order in the field must take the form \cite{chaikin}
    \begin{equation}\label{Eq:Goldstonefreeenergy}
        F =  \int d^{2}x \; \left[ \frac{K}{2} ( \partial_{i} \phi^{i} )^2 + \frac{G}{2} \left( \partial_{i} \phi_{j} \partial^{i} \phi^{j} + k_{0}^2 \phi_{i} \phi^{i} \right)  + \theta_{1} \partial_{i} \phi^{i} + \theta_{2} \epsilon^{ij} \partial_{i} \phi_{j} \right] 
    \end{equation}
where $K$ is the bulk modulus, $G$ the shear modulus and $k_{0}$ a small mass term for the Goldstone boson, which takes into account a possible small explicit component of translational symmetry breaking. In writing $F$ we have made use of spatial rotational invariance, and the square symmetry of the underlying lattice \cite{Delacretaz:2017zxd}. We have included $\theta_{1}$ and $\theta_{2}$ as the sources for the pseudo-Goldstone bosons, corresponding to the operators $\lambda_{1} = \partial_{i} \phi^{i}$ and $\lambda_{2} = \epsilon^{ij} \partial_{i} \phi_{j}$, that will enter our hydrodynamic description. 

The form of free energy $F$ implies the Poisson bracket
    \begin{eqnarray}
        \left\{  \phi_{i}(t,\vec{x}), \pi^{j}_{\phi}(t,\vec{y}) \right\}
        &=& \left( \delta\indices{_{i}^{j}} + \partial^{j} \phi_{i}(t,\vec{x}) \right) \delta^{2}(\vec{x}-\vec{y}) 
    \end{eqnarray}
where $\pi^{i}_{\phi}$ is the translational Noether charge of the pseudo-Goldstone bosons. Of the two terms on the right hand side the latter is the usual one for translating a vector field by a constant shift. The former however indicates that the bosons have a non-zero ``charge'' under translations. This is important as it subsequently implies non-conservation of the spatial momentum for our model
    \begin{eqnarray}
          \label{Eq:HamiltonianEvolutionBosons}
          \partial_{t} \pi^{i}
      &=& \left\{ \pi^{i}, H \right\} 
      = - G k_{0}^2 \phi^{i} + \mathrm{higher \; derivatives}
    \end{eqnarray}
where $H$ is the Hamiltonian of the full theory. Physically, even in the absence of other dissipative effects, because the boson is not massless ($k_{0} \neq 0$) our system will lose momentum. Indeed, the appearance of these terms \eqref{Eq:HamiltonianEvolutionBosons} in the relaxed conservation equation of the momentum are responsible for a non-zero pinning frequency.

Moreover, the contribution of $\phi_{i}$ to the spatial stress tensor can be obtained from $F$ by applying the usual Nother procedure:
    \begin{eqnarray}
      \label{Eq:NoetherSEM}
      T_{ij}^{\phi} &=& - (G+K) \partial_{k} \phi^{k} \delta_{ij} - G (\partial_{i} \phi_{j}-\delta_{ij} \partial_{k} \phi^{k}) \; .
    \end{eqnarray}
This expression will appear at leading order in our constitutive relation for the spatial stress tensor of the full theory ($T^{ij}$).

Finally, using the static equations of motion derived from \eqref{Eq:Goldstonefreeenergy}, and Fourier transforming to momentum space with momentum $\vec{k}=\{k_x, k_y \}$, one can compute the static susceptibilities associated to the operators $\lambda_1$ and $\lambda_2$:
\begin{equation}
\chi_1=\partial_{\theta_1}\lambda_1=\frac{\vec{k}^2}{G k_{0}^2 + (K + G) \vec{k}^2 } \ , \qquad \chi_2=\partial_{\theta_2}\lambda_2=\frac{\vec{k}^2}{G ( k_{0}^2 + \vec{k}^2 )} \ .
\end{equation}
Given these considerations, the static susceptibility matrix for our theory will have the form
    \begin{eqnarray}
        \label{Eq:StaticSusceptibility}
     \chi &=& \left(
        \begin{array}{cccccc}
          \partial_{T} s & \partial_{\mu} s & 0 & 0 & 0 & 0 \\
          \partial_{T} n & \partial_{\mu} n & 0 & 0 & 0 & 0 \\
          0 & 0 & \chi_{\pi \pi} & 0 & 0 & 0 \\
          0 & 0 & 0 & \chi_{\pi \pi} & 0 & 0 \\
          0 & 0 & 0 & 0 & \chi_{1}  & 0 \\
          0 & 0 & 0 & 0 & 0 & \chi_{2}  \\
        \end{array}
     \right) \; . 
    \end{eqnarray}
where $\chi_{\pi \pi}$ is the momentum susceptibility, $s$ the entropy density, $n$ the charge density, $\pi^{i}$ the momentum density and we work in the basis of sources $s_{A}=(T, \mu,v_x,v_y, \theta_1,\theta_2)$ with $T$ the temperature, $\mu$ the chemical potential and $\vec{v}=\{v_x,v_y\}$ the spatial velocity of the fluid. 

 With the susceptibilities to hand, one can now perform the desired perturbation analysis about a state of constant temperature ($T + \delta T$) and chemical potential ($\mu + \delta \mu$) with all other background source fields, in particular sources for the pseudo-Goldstone bosons and spatial velocity, vanishing. The system is then described in global thermodynamic equilibrium by an equation of state $p(T,\mu)$, with $p$ being the pressure of the system, whose exact form is, for the time being, unimportant. Additionally we will allow for a non-zero background magnetic field $F^{ij} = B \epsilon^{ij}$ which is globally small and constant \footnote{We define $\epsilon^{12}=1$.} and assume that whatever microscopic theory faithfully describes our system has spatial parity invariance \footnote{In two spatial dimensions the parity operation is defined to act on only one of the Cartesian coordinates e.g.~$(x,y) \mapsto (x,-y)$.}.

Electric charge and the entropy density will be conserved in our description (the latter because we work only to first order in derivatives), while  the momentum density and the boson fields will be allowed to relax in their (non-)conservation equations i.e.~
  \begin{eqnarray}
  \label{Eq:conservation}
    & \partial_{t}& \left( n, s \right) + \partial_{i} \left( J^{i}, Q^{i} / T \right) = 0 \; , \\
    \label{Eq:RelaxedMomentum}
    &\partial_{t}& \pi^{i} + \partial_{j} T^{ji} = F^{ij}J_j- \Gamma^{ij}\pi_j-k_o^2 G \phi^i \; , \\
    \label{Eq:JosephsonEquation}
    &\partial_{t}& \lambda_{a} + \partial_{i} J_{a}^{i} = -\Omega\indices{_{a}^{b}} \lambda_{b} =- \Omega_{1} \lambda_{a} - B \Omega_{2}  \epsilon_{ab} \lambda^{b}  \; ,
  \end{eqnarray}
where $J^{i},Q^{i},T^{ij},J_{a}^{i}$ are the charge density and heat current, the spatial stress tensor and the Goldstone current respectively. The first term on the right hand side of \eqref{Eq:RelaxedMomentum} is simply the expression of the Lorentz force, while $\Gamma^{ij}$ is a momentum relaxation matrix accounting for microscopic contributions to momentum loss. We shall make the assumption that $\Gamma^{ij} =\Gamma \delta^{ij}$ so that microscopic dissipative processes only relax the momentum parallel to the fluid velocity. Relaxation perpendicular to $v^{i}$ therefore only occurs through the Lorentz term and is proportional to the magnetic field. The last term in \eqref{Eq:RelaxedMomentum} is a consequence of the evolution equation  \eqref{Eq:HamiltonianEvolutionBosons}. Finally, equation \eqref{Eq:JosephsonEquation} is often referred to as the Josephson condition in analogy with the hydrodynamic description of superfluids. There, $\Omega\indices{_{a}^{b}}$ is a relaxation rate matrix for the pseudo-Goldstone modes, and it has been decomposed in the longitudinal and Hall contributions $\Omega_1$ and $\Omega_2$ employing spatial isotropy with respect to the square crystal symmetry. As proven in \cite{Amoretti:2018tzw,Amoretti:2019cef}, the relaxation matrix $\Omega\indices{_{a}^{b}}$ is nontrivial as soon as the breaking of translations is pseudo-spontaneous.

 The expressions defining the currents $J_{A} = (Q^{i}/T,J^{i},T^{ij},J^{i}_{a})$ in terms of the source fields are the aforementioned constitutive relations. To determine them one should in principle work order by order in derivatives, adding at each order all possible tensor structures that multiply derivatives of the sources while respecting any symmetries. For our purposes, these constitutive relations can be linearised around their background values to have the form
  \begin{eqnarray}
   \label{Eq:HeatCurrent}
   \frac{Q^{i}}{T} &=& s v^{i} - \alpha_{0} \left( \partial^{i} \delta \mu - F^{ij} v_{j} \right) - \frac{\bar{\kappa}_{0}}{T} \partial^{i} \delta T - \gamma_{2} \partial^{i} \theta_{1} \; , \\
   \label{Eq:ChargeCurrent}
   J^{i} &=& n v^{i} - \sigma_{0} \left( \partial^{i} \delta \mu - F^{ij} v_{j} \right) - \alpha_{0} \partial^{i} \delta T - \gamma_{1} \partial^{i} \theta_{1} \; , \\
   \label{Eq:SpatialSEM}
   T^{ij} &=& \left( n \delta \mu + s \delta T - \left( G + K \right) \chi_{1} \theta_{1} \right) \delta^{ij} - G \chi_{2} \theta_{2}  \epsilon^{ij} \nonumber \\
   &\;& - \eta \left( \partial^{i} v^{j} + \partial^{j} v^{i} - \partial_{k} v^{k} \delta^{ij} \right) - \zeta \partial_{k} v^{k} \delta^{ij} + \gamma_{1} B \theta_{2} \delta^{ij} \; , \\
   \label{Eq:GoldstoneVector1}
   J_{1}^{i} &=& - v^{i} - \gamma_{1} \left( \partial^{i} \delta \mu - F^{ij} v_{j} \right) - \gamma_{2} \partial^{i} \delta T - \xi_{1} \chi_{1} \partial^{i} \theta_{1} + \xi_{2} \chi_{2} \epsilon^{ij} \partial_{j} \theta_{2} \; , \\
   \label{Eq:GoldstoneVector2}
   J_{2}^{i} &=& \epsilon\indices{^{i}_{j}} J_{1}^{j} \; , 
  \end{eqnarray}
where $\theta_{1}$, $\theta_{2}$ and $\vec{v}$ are all fluctuations about vanishing background values.  The terms in \eqref{Eq:HeatCurrent}-\eqref{Eq:GoldstoneVector2} have been constrained such that the system has time reversal covariance, spatial parity covariance and respects Onsager reciprocity. Consequently, we set the incoherent Hall conductivities $(\sigma_{xy}^{0},\alpha_{xy}^{0},\bar{\kappa}_{xy}^{0})$ to zero; when $B=0$ this follows from spatial parity invariance. Order $B$ corrections to the incoherent Hall conductivities would necessarily be order $O(\partial^2)$ in the hydrodynamic expansion and therefore will not have any relevance to our current truncation of the constitutive relations. Moreover, we expect such corrections could only arise if the fixed point was not independent of $B$ at first order, which would be incompatible with treating the magnetic field as external. 

The displayed transport coefficients of \eqref{Eq:conservation}-\eqref{Eq:JosephsonEquation}  and \eqref{Eq:HeatCurrent}-\eqref{Eq:GoldstoneVector2}, (namely $\Gamma$, $\Omega_1$, $\Omega_2$, $\sigma_0$, $\alpha_0$, $\bar{\kappa}_0$, $\eta$, $\zeta$, $\gamma_1$, $\gamma_2$,  $\xi_1$, and $\xi_2$), are arbitrary functions of $T$ and $\mu$ in the framework of hydrodynamics. Their exact dependence on the temperature and chemical potential can in principle be determined from a microscopic theory for the system. However to determine the form of the AC transport coefficients we do not require such information (which is the key strength of the hydrodynamic approach). Indeed, the displayed coefficients are only constrained by imposing the positivity of entropy production, namely
 \begin{equation}
 \partial_t s+\partial_i \left( \frac{Q^i}{T} \right) \ge 0 \ .
 \end{equation}
This latter requirement leads to bounds on particular transport coefficients including those typical for a charged fluid with momentum loss
  \begin{eqnarray}
    \sigma_{0}, \; \bar{\kappa}_{0}, \; \eta \; , \Gamma \; , \Omega_{1} \geq 0 \; , \qquad \bar{\kappa}_{0} \sigma_{0} - T \alpha_{0}^2 \geq 0 \; , \qquad 
  \end{eqnarray}
which can be derived by examining the retarded charge-charge Green's function and imposing that its imaginary part is positive definite \cite{Kovtun:2012rj,Delacretaz:2019wzh}. For the pseudo-Goldstone boson we additionally have the following useful constraint: $\xi_{1} > 0$. This subsequently leads to bounds on $\gamma_{1}$ and  $\gamma_{2}$:
  \begin{eqnarray}\label{Eq:boundgamma}
    (\gamma_{1}^2 ,\gamma_{2}^2)\leq \left(\sigma_{0},\frac{\bar{\kappa}_{0}}{T}\right) \mathrm{min}\left[ \frac{\xi_{1}}{K+G} , \frac{\Omega_{1}}{\chi_{\pi \pi} \omega_{0}^2} \right] \; . 
  \end{eqnarray}
Consequently we can assume that $\gamma_{1}$ and $\gamma_{2}$ are small compared to other scales in our problem.

Fourier transforming to momentum space, with $\vec{k}=\{k_x,k_y\}$, the conservation laws \eqref{Eq:conservation}-\eqref{Eq:JosephsonEquation}, together with the constitutive relations \eqref{Eq:HeatCurrent}-\eqref{Eq:GoldstoneVector2}, can be organized in the following way:
  \begin{eqnarray}
    \label{Eq:PerturbedFluidSystem}
    \partial_{t} q_{A}(t,\vec{k}) + M\indices{_{A}^{C}}(\vec{k},B) s_{C}(t,\vec{k}) &=& 0
  \end{eqnarray}
where $q_{A}=(s,n,\pi^{i},\lambda_{a})$ are the charges and $s_{A}=(\delta T, \ \delta \mu, \ v_i, \ \theta_a)$ are the linearized sources. The matrix $M_{AC}$ is defined as follows: {\scriptsize
    \begin{eqnarray}
     \left(
        \begin{array}{cccccc}
         \frac{\bar{\kappa}_0}{T} \vec{k}^2 & \alpha_0 \vec{k}^2 & i s k_{x} - i \alpha_0 B k_{y}  & i s k_{y}+i \alpha_0 B k_{x} & \gamma_{2} \vec{k}^2 & 0 \\
         \alpha_0 \vec{k}^2 & \sigma_0 \vec{k}^2 & i n k_{x}- i \sigma_0 B k_{y} & i  n k_{y}+i \sigma_0 B k_{x} & \gamma_{1} \vec{k}^2 & 0 \\
         i s k_{x}+ i \alpha_0 B k_{y} & i n k_{x}+i \sigma_0 B k_{y} & \Gamma \chi_{\pi \pi} + \sigma_0 B^2 + \eta \vec{k}^2+\zeta k_x^2 & \zeta k_{x} k_{y} -n B & - i k_{x}+i \gamma_{1} B k_{y} & i k_{y}+i \gamma_{1} B k_{x} \\
         i s k_{y}- i \alpha_0 B k_{x} & i n k_{y} - i \sigma_0 k_{x} & \zeta k_{x} k_{y}+n B & \Gamma \chi_{\pi \pi} + \sigma_0 B^2 + \eta \vec{k}^2+\zeta k_{y}^2  & - i k_{y}-i \gamma_{1} B k_{x} & - i k_{x}+i \gamma_{1} B k_{y} \\
         \gamma_{2} \vec{k}^2 & \gamma_{1} \vec{k}^2 & - i k_{x}- i \gamma_{1} B k_{y} & - i k_{y}+i \gamma_{1} B k_{x} & \left( \Omega_{1} + \xi_{1} \vec{k}^2 \right) \chi_{1} & \Omega_{2} B \chi_{2} \\
         0 & 0 & i k_{y}- i \gamma_{1} B k_{x} & - i k_{x}- i \gamma_{1} B k_{y} & - \Omega_{2} B \chi_{1} & \left( \Omega_{1} + \xi_{2} \vec{k}^2 \right) \chi_{2}
        \end{array}
     \right) \ . \nonumber
    \end{eqnarray}
}

Given $M_{AB}$, one can straightforwardly apply the Martin-Kadanoff procedure \cite{kadanoff1963} to obtain the retarded Green's functions for the charges:
    \begin{eqnarray}\label{Eq:retardedgr}
     \langle q_{A}(\omega,\vec{k}) q_{B}(-\omega,-\vec{k}) \rangle_{R} &=& - \left( \mathbbm{1}_{6} + i \omega \left( - i \omega \mathbbm{1}_{6} + M \chi^{-1} \right)^{-1} \right) \chi  \; .
    \end{eqnarray}
The transport coefficients are related to \eqref{Eq:retardedgr} using the Kubo formulae (see e.g.~\cite{Kovtun:2012rj}). In this paper we are interested in analyzing the transport coefficients in the limit $\gamma_1, \ \gamma_2, \ \Omega_2 \sim 0$. The assumption on $\gamma_{1,2}$ comes from the bound \eqref{Eq:boundgamma}, while $\Omega_2$ is subleading with respect to $\Omega_1$ in the low magnetic field approximation. With these assumptions the electric, thermo-electric and thermal conductivities $\hat{\sigma}=\sigma_{xx}\mathbbm{1}_{2}+\sigma_{xy} \hat{\epsilon}$, $\hat{\alpha}=\alpha_{xx}\mathbbm{1}_{2}+\alpha_{xy} \hat{\epsilon}$ and $\hat{\bar{\kappa}}=\bar{\kappa}_{xx}\mathbbm{1}_{2}+\bar{\kappa}_{xy} \hat{\epsilon}$ assume the following form: 
  \begin{eqnarray}\label{Eq:finaltc}
     \hat{\Sigma} &=& \hat{\Sigma}_{0} +\frac{\left( - i \omega  + \Omega_1  \right)}{\chi_{\pi \pi}} \hat{C}^{-1} \cdot A_{\Sigma} 
  \end{eqnarray}
where $\hat{\Sigma}=(\hat{\sigma},\hat{\alpha},\hat{\bar{\kappa}})$, $\hat{\Sigma}_{0}=(\sigma_{0},\alpha_{0},\bar{\kappa}_{0}) \mathbbm{1}_{2}$, $ \hat{A}_{\left(\sigma,\alpha,\bar{\kappa}\right)} = \left( \hat{q} \cdot \hat{q}, \hat{q} \cdot \hat{S} , T \hat{S} \cdot \hat{S} \right)$ and 
  \begin{equation}
    \hat{S} = s \mathbbm{1}_2 + \alpha_{0} B \hat{\epsilon} \; , \qquad \hat{q}=n\mathbbm{1}_2+\sigma_0B \hat{\epsilon} \ , \qquad C=(\Omega_1-i \omega)\left((\Gamma+\gamma_c)\mathbbm{1}_2-\omega_c \hat{\epsilon}\right)+\omega_0^2 \mathbbm{1}_2 \ .
  \end{equation}
$\omega_0=k_0 \sqrt{\frac{G}{\chi_{\pi \pi}}}$ is the pinning frequency of the CDW, while $\omega_c=\frac{n B}{\chi_{\pi \pi}}$ and $\gamma_c=\frac{\sigma_0B^2}{\chi_{\pi \pi}}$ are the cyclotron frequency and the cyclotron mode respectively \cite{Hartnoll:2007ih}. The electric resistivity matrix is defined as the inverse of $\hat{\sigma}$, $\hat{\rho}\equiv\hat{\sigma}^{-1}$, while the quantity $\hat{\bar{\kappa}}$ appearing in \eqref{Eq:finaltc} is the thermal conductivity at zero electric field. This latter quantity is related to the thermal conductivity at zero electric current which can be determined by experimental measurements through the relation $\hat{\kappa}=\hat{\bar{\kappa}}-T \hat{\alpha}\cdot \hat{\rho} \cdot\hat{\alpha}$. Finally, the Nernst coefficient $N$ is defined as the $xy$ component of the matrix $\hat{\rho} \cdot \hat{\alpha}$.

By taking the DC limit $\omega \rightarrow0$ of \eqref{Eq:finaltc}, and expanding at low magnetic field $B$, one eventually finds the transport coefficients outlined in the main text (equations (2) in the body of our paper).

Finally we note that the coefficients $\sigma_0$, $\alpha_0$, $\bar{\kappa}_0$, $ \gamma_1$ and $\gamma_2$ are further constrained by imposing relativistic covariance of the equations \eqref{Eq:conservation}-\eqref{Eq:GoldstoneVector2}; meaning that even though the ground state of the system breaks relativistic invariance, the fluctuation equations are covariant under the Lorentz group. This is a typical situation happening in the vicinity of a quantum critical point with emergent relativistic invariance \cite{sachdev2011}. In particular, relativistic covariance imposes that the heat current $Q^i$ must be a function of the momentum density and the electric current, namely $Q^i=\pi^i-\mu J^i$. This implies:
\begin{equation}
\alpha_0=-\frac{\mu \sigma_0}{T}, \qquad \bar{\kappa}_0=\frac{\mu^2 \sigma_0}{T}, \qquad \gamma_2=-\frac{\mu \gamma_1}{T}.
\end{equation}

\section{Experimental details}
\label{expmet}
\renewcommand{\thefigure}{A\arabic{figure}}
\setcounter{figure}{0}
{\bf Growth and characterization of samples} The details about the synthesis of the {Bi}$_2${Sr}$_2${CuO}$_6$ single crystals are reported in \cite{piriou2010}. The quality of the crystal was checked by x-ray diffraction (XRD) and energy-dispersive x-ray microprobe (EDX). The XRD pattern measured in a Bragg-Brentano geometry is reported in \cite{piriou2010} and XRD data prove the high crystallinity of the samples. The chemical composition for the crystals was checked by EDX. The average composition measured over large crystal areas is {Bi}$_{2.05}${Sr}$_{1.98}${Cu}$_{0.98}${O}$_{6.04}$ with errors on the local deviations in formula units of $\Delta$(Bi)=0.05, $\Delta$(Sr)=0.05, $\Delta$(Cu)=0.02. 
The crystals were cleaved from the core of a precursor rod with typical lengths of 1-5 mm and thickness of 0.1-0.2 mm and they are superconducting with an onset of the magnetic susceptibility transition at 11 K and a transition width of about 4 K (see below).\\

{\bf Electrical measurements} The electrical measurements have been performed using a commercial system \textit{Physical Property Measurement System} (PPMS) by Quantum Design in a range of temperature between 2 and 400 K and in magnetic fields from -9 to 9 T. High-quality contacts of low resistance (tens of Ohm) were made by using an Ag paste which was cured in air at about 250$^{\circ}$C. The electric resistivity was measured using a standard 4-probe technique, whereas we used a standard 6-terminal method for simultaneous measurements of the longitudinal and transverse voltage. Data were taken by sweeping $B$ from -9 T to 9 T at fixed temperatures, in order to separate the even (magnetoresistance) and odd (Hall resistivity) components of the signal with respect to $B$. 

\begin{figure}
	\centering
	\includegraphics[width=0.8\linewidth]{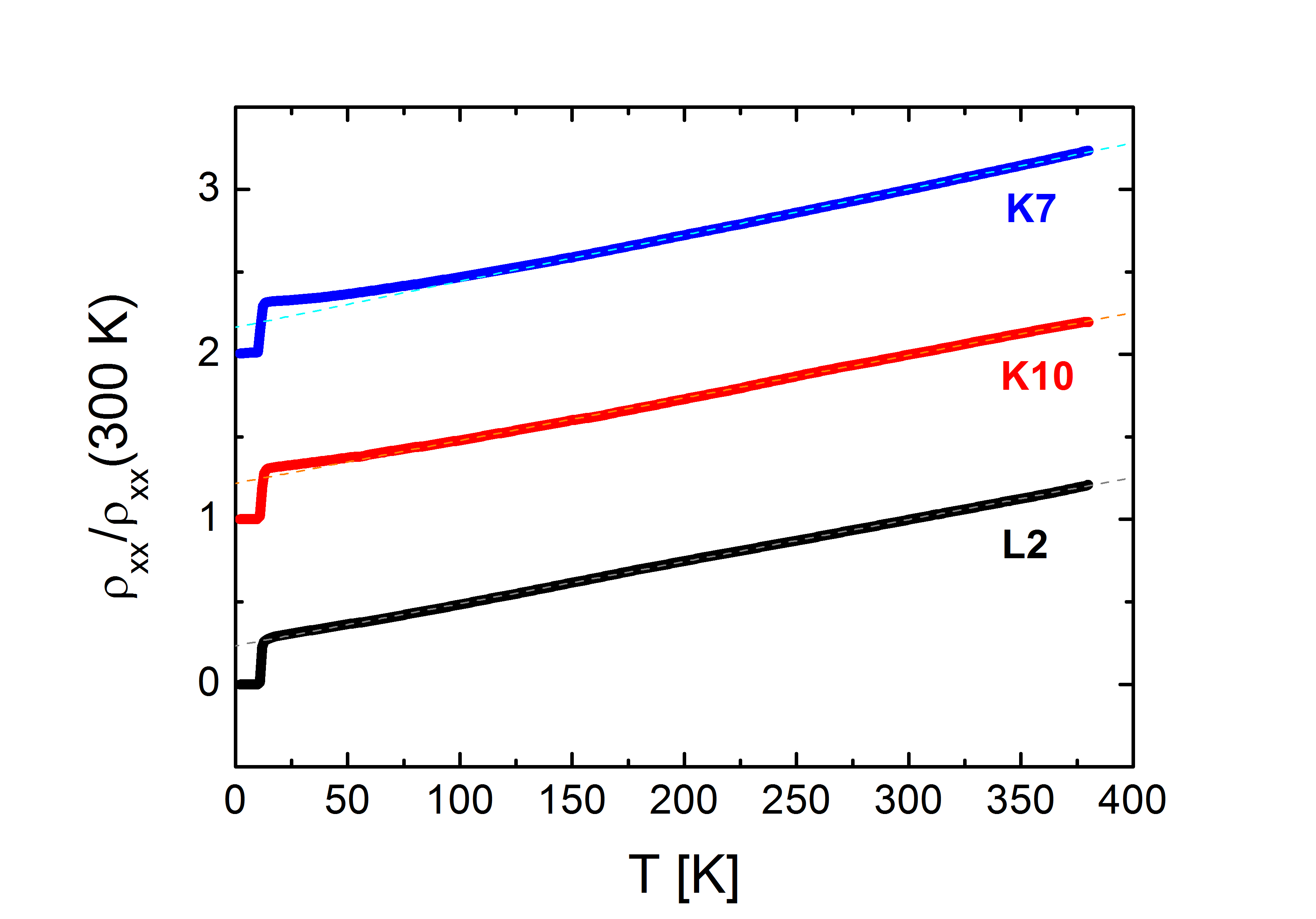}
	\caption{Temperature dependence of $\rho_{xx}$  normalized to  $\rho_{xx}(300 \; \mathrm{K})$}
	\label{RhoSM}
\end{figure}

The electric characterization has been performed on several samples and we report here the data relative to three crystals, named L2, K8 and K10.
Figure \ref{RhoSM} shows the in-plane electric resistivity $\rho_{xx}$ of the L2, K8 and K10 samples as a function of temperature from 4 to 380 K. The data are normalized to the value at 300 K ($\rho_{xx}/\rho_{xx}(300 \; \mathrm{K})$) and the curves are shifted to avoid overlapping. The crystals show a quite sharp superconducting transition which occurs at $T_c\sim 11$~K. Above this temperature $\rho_{xx}(T)$ is linear up to 380~K. Deviations from linearity can be observed in K7 and K10 samples at low temperatures and this behavior may be ascribed to a different concentration of either oxygen or impurities in these samples.

\begin{figure}[h]
	\centering	
	\begin{minipage}{.51\textwidth}
		\centering
		\includegraphics[width=1\linewidth]{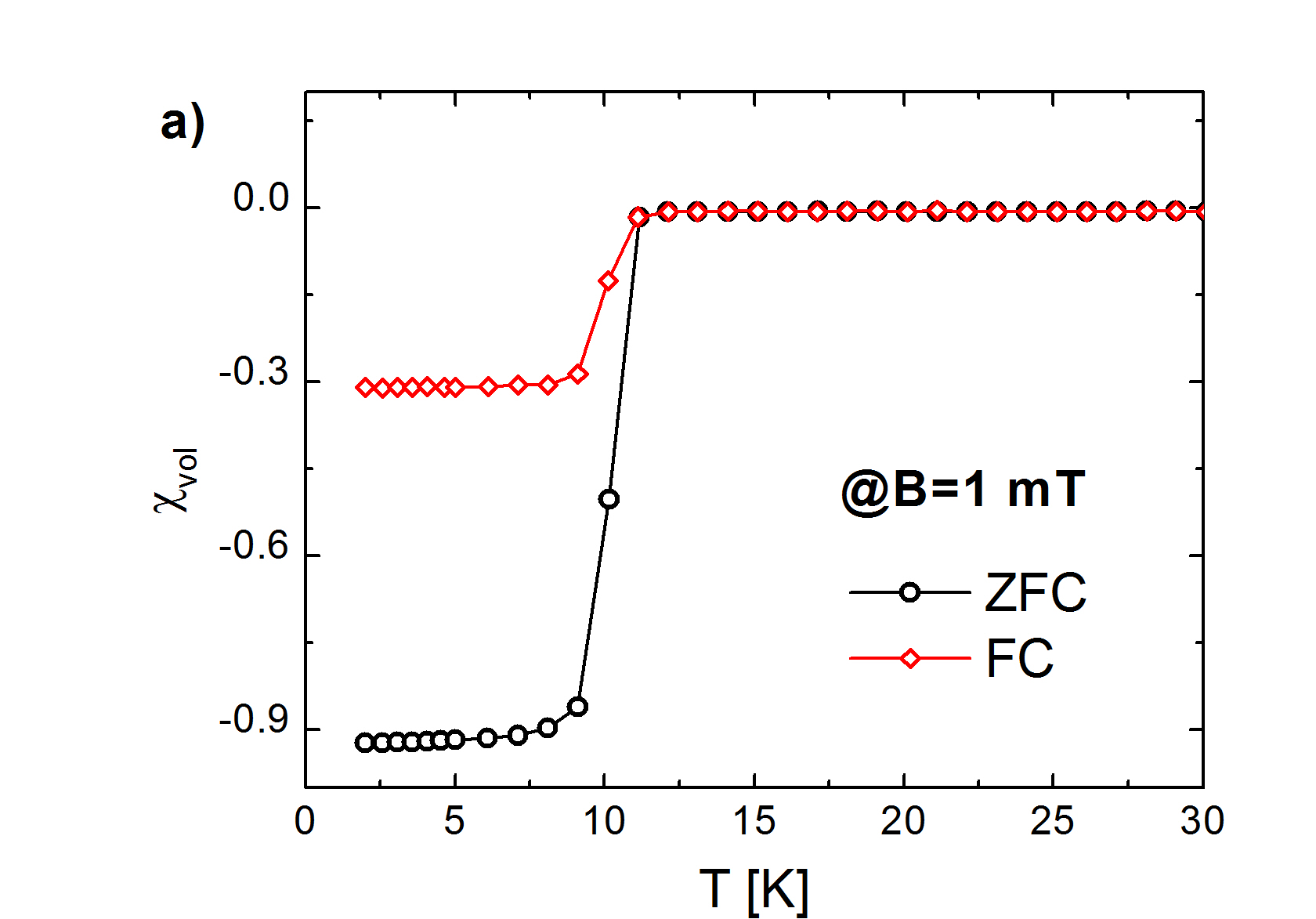}
	\end{minipage}%
	\begin{minipage}{.5\textwidth}
		\centering
		\includegraphics[width=1\linewidth]{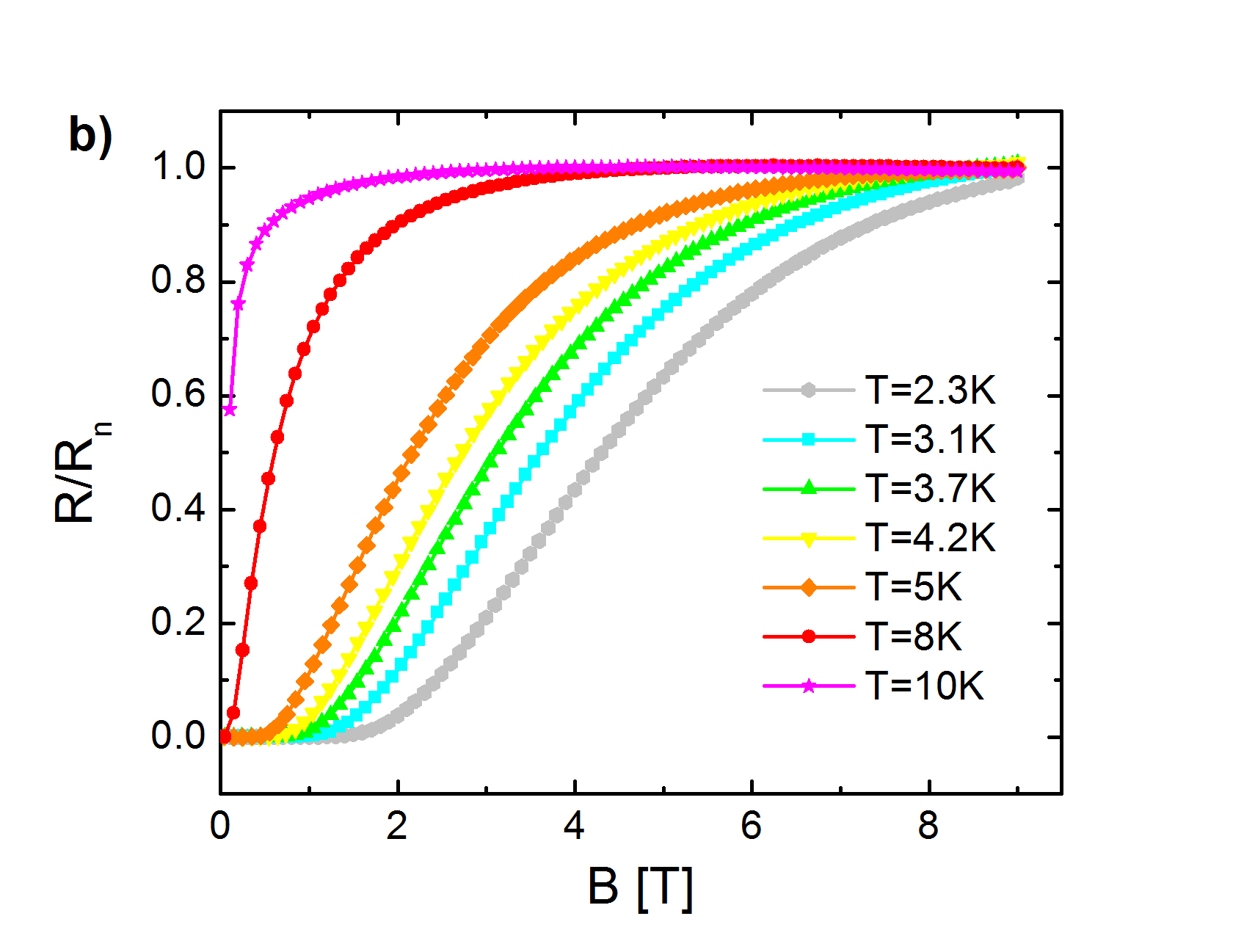}
	\end{minipage}%
	\caption{{\bf a$)$} Temperature dependence of the magnetic susceptibility $\chi_{vol}$ between 2 and 30 K in an applied magnetic field $B$=1 mT. Black empty dots represent the zero-field cooled (ZFC) measurement, whereas red empty bullets show the field-cooled (FC) one. {\bf b$)$} Electrical resistance as a function of magnetic field from 0 to 9 T at selected temperatures in the range 2.3-10 K. The curves are normalized to the normal state value of the electrical resistivity $R_{n}$.}
	\label{figA1}
\end{figure}

Figure \ref{figA1}b shows the $B$-dependence of the electrical resistance $R$ of the L2 sample as a function of $B$ from 0 to 9 T at selected temperatures below $T_c$ from 2.3 to 10 K. The curves are normalized to the normal state value of the electric resistance $R_{n}$ evaluated at $T=10$ K and $B=9$ T. It can be noted that at $T=10$ K, $R$ saturates to $R_n$ for $B\geq 2$ T and, with lowering $T$, the normal state is reached at higher and higher fields in accordance to what is expected for a type-II superconductor (at $T=2.3$ K $R_n$ is reached at 9 T).\\

{\bf Magnetic susceptibility measurement} The temperature dependence of the magnetic susceptibility $\chi_{vol}$ of the L2 sample was measured by a commercial dc-SQUID magnetometer (MPMS2 by Quantum Design). Figure \ref{figA1}a shows $\chi_{vol}$ as a function of temperature between 2 and 30 K in an applied magnetic field $\mu_0B$= 1 mT, parallel to the $c$ axis of the crystal. The sample was first cooled down to 2 K, then $B$ was switched on and the sample heated up to 30 K (ZFC curve, black empty bullets in figure \ref{figA1}a). The sample was then cooled down to 2 K in field and measurement repeated by heating the sample up to 30 K (FC curve, red empty bullets). The superconducting transition is narrow, with an onset temperature $T_c\sim$11 K and the saturation value $\chi_{vol}\sim$-0.9 is reached around 8 K. This is a proof of the homogeneity of the crystal.
The low temperature saturation value $\chi_{vol}\sim$-0.9 corresponds to a 90\% Meissner fraction which implies that almost all the sample is superconducting. This value has been evaluated taking into account a demagnetization factor $N_{\parallel}\sim$0.95, considering that $B$ is parallel to the $c-$axis of the crystal \cite{poole1995}. \\

{\bf Thermo-electrical and thermal measurements} The thermo-electric and thermal characterizations were performed using a liquid-He cryostat from Oxford Instruments endowed with 15/17 T superconducting magnet and home-made probe, in a temperature range 4-300 K and in magnetic fields from -14 T to 14 T. The thermal circuit is realized by gluing the bottom of the sample to the copper sample stage of the probe using a tested electrical insulating glue with high performance in heat conduction. A 3 K$\Omega$-resistive heater is glued on the top of the sample to apply a heating power $W_H$ which creates a longitudinal temperature gradient across the sample ($\Delta T_x$). By applying a magnetic field $B$ along the $c$-axis of the crystal (along $z$), normal to the {CuO}$_2$ planes, one creates both a transverse temperature gradient ($\Delta T_y$) and the Nernst signal ($\Delta$V$_y$). The temperature gradient was estimated using Chromel-Au differential thermocouple, whose tips were glued to the sample. The thermocouples have been calibrated up to 16 T in a range of temperatures between 4 and 350 K. The electrodes to pick up the Nernst signal consist of calibrated copper wires attached to the sample through Ag paste which was cured in air at about 250$^{\circ}$C. We performed all the thermal and thermoelectric measurements with both positive and negative $B$ in order to separate the even (Seebeck effect and longitudinal thermal conductivity) and odd (Nernst effect and transverse thermal conductivity) parts of the signal with respect to $B$. All the measurements have been carried out in conditions of temperature and magnetic field stability. The Nernst coefficient is defined as $N$=($\Delta$V$_y$/$l_N$)($\Delta T_x$/$l_x$), the longitudinal thermal conductivity as $\kappa_{xx}$=($W_H$/$\Delta T_x$)($l_x/wt$) and the thermal Hall conductivity as $\kappa_{xy}$= $\kappa_{xx}$($\Delta T_y$/$\Delta T_x$)($l_x/w$), where $l_N$, $l_x$, $w$,  and $t$ are respectively the distance between the Nernst electrodes, the distance between the longitudinal thermocouple tips, the width of the sample (along $y$) and its thickness (along $z$).
To estimate $\kappa_{xy}$ as a function of temperature (figure 2c of main text), we considered the slope $\Delta$($\Delta T_y$/$w$)/$\Delta B$, so that $\kappa_{xy}$=$\frac{\Delta(\Delta T_y/w)}{\Delta B} B \frac{\kappa_{xx}}{\Delta T_x/l_x}$.

Typically $\Delta T_x/l_x$ and $\Delta T_y/w$ were of the order of magnitude of (0.1-1)$\times$10$^3$ K/m and (0.1-1) K/m respectively. The ratio $\frac{k_{xx}}{\Delta T_x/l_x}$ has been measured during a previous longitudinal thermal conductivity measurement on the same crystal. The error on the magnitude of $\kappa_{xy}$ (error bars in the inset of figure 2c) has been estimated as the root mean square error of the $\Delta T_y/w$ data and it results between $\pm$20 and $\pm$30\%.\\

\begin{figure}
\centering
\includegraphics[width=0.8\linewidth]{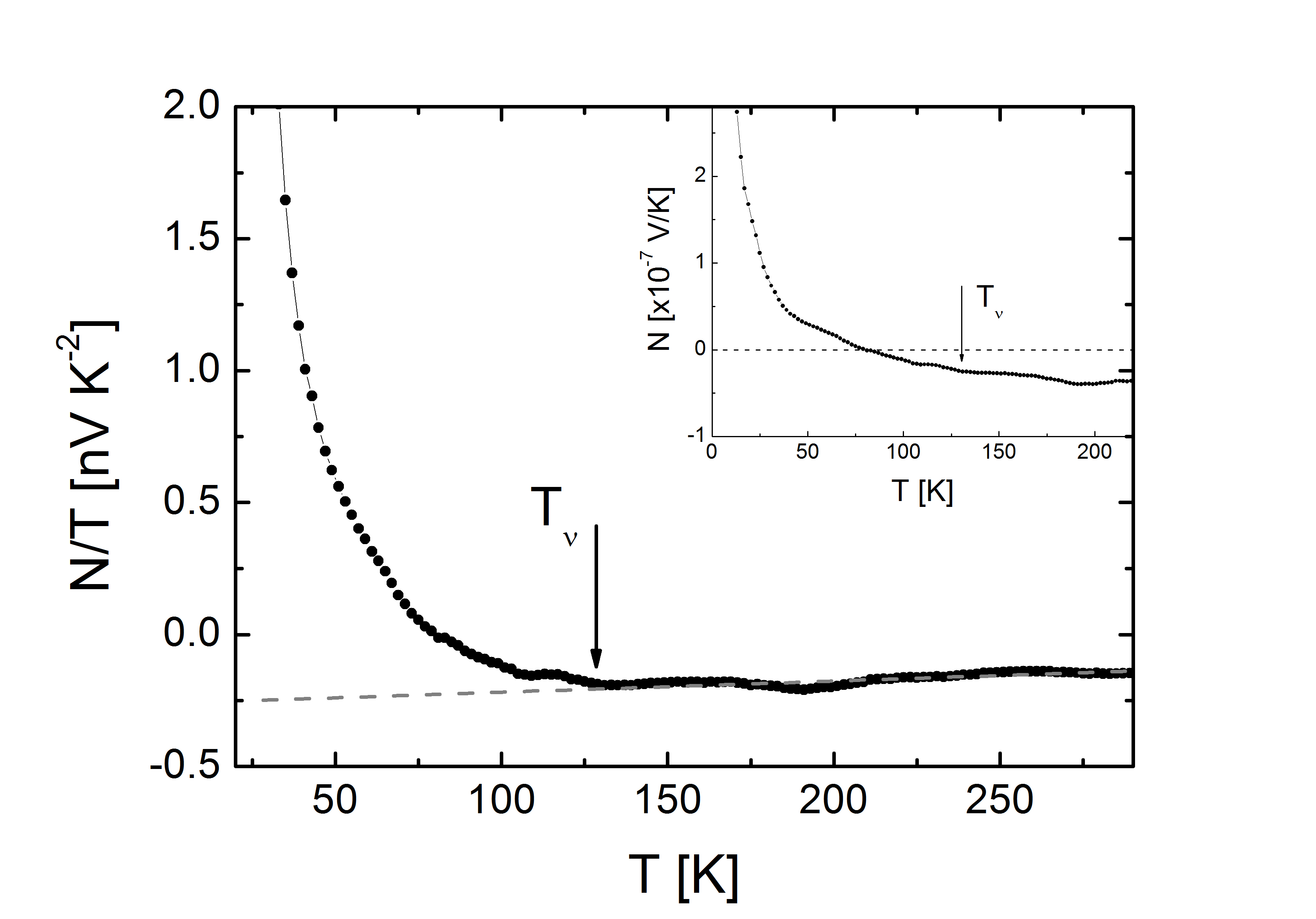}
\caption{Nernst coefficient $N$ divided by temperature $T$ in the temperature range 20-180 K. The measurements are performed at $B=$14 T. The onset temperature $T_\nu$ (black arrow) is defined as the deviation of $N/T$ from a linear fit at high temperature (grey dashed line). We therefore estimate $T_N=130$ K $\pm 10$ K. Inset: $N$ between 0 and 220 K. The black arrow indicates $T_\nu$.}
\label{NoverT}
\end{figure}

{\bf Estimation of $T_\nu$} We define $T_\nu$ as the point where $N/T$ deviates from linearity at high temperature (see Figure \ref{NoverT}). This is the same criterion as the one proposed in ref. \cite{taillefer2009} and it is based on the fact that $N/T$ is linear in $T$ in both Nd- and Eu-LSCO whenever there is neither superconducting contribution to $N$ nor any stripe/charge ordering \cite{grissonnanche2019,Hess2010}. This qualitative definition allows to estimate $T_\nu$ within $\pm 10$ K.\\

{\bf Specific heat measurement} Low temperature specific heat measurements were performed using a Quantum Design PPMS by means of the two-tau method, both in zero and 7 T magnetic field. A mosaic of five crystals with $c$-axes parallel to the magnetic field has been assembled to reach a total mass of 4.5 mg. The heat capacity of the addenda (i.e. sample holder and 0.5 mg of Apiezon N) have been previously measured and subtracted from the total heat capacity.

\begin{figure}
	\centering
	\begin{minipage}{.5\textwidth}
		\centering
		\includegraphics[width=1\linewidth]{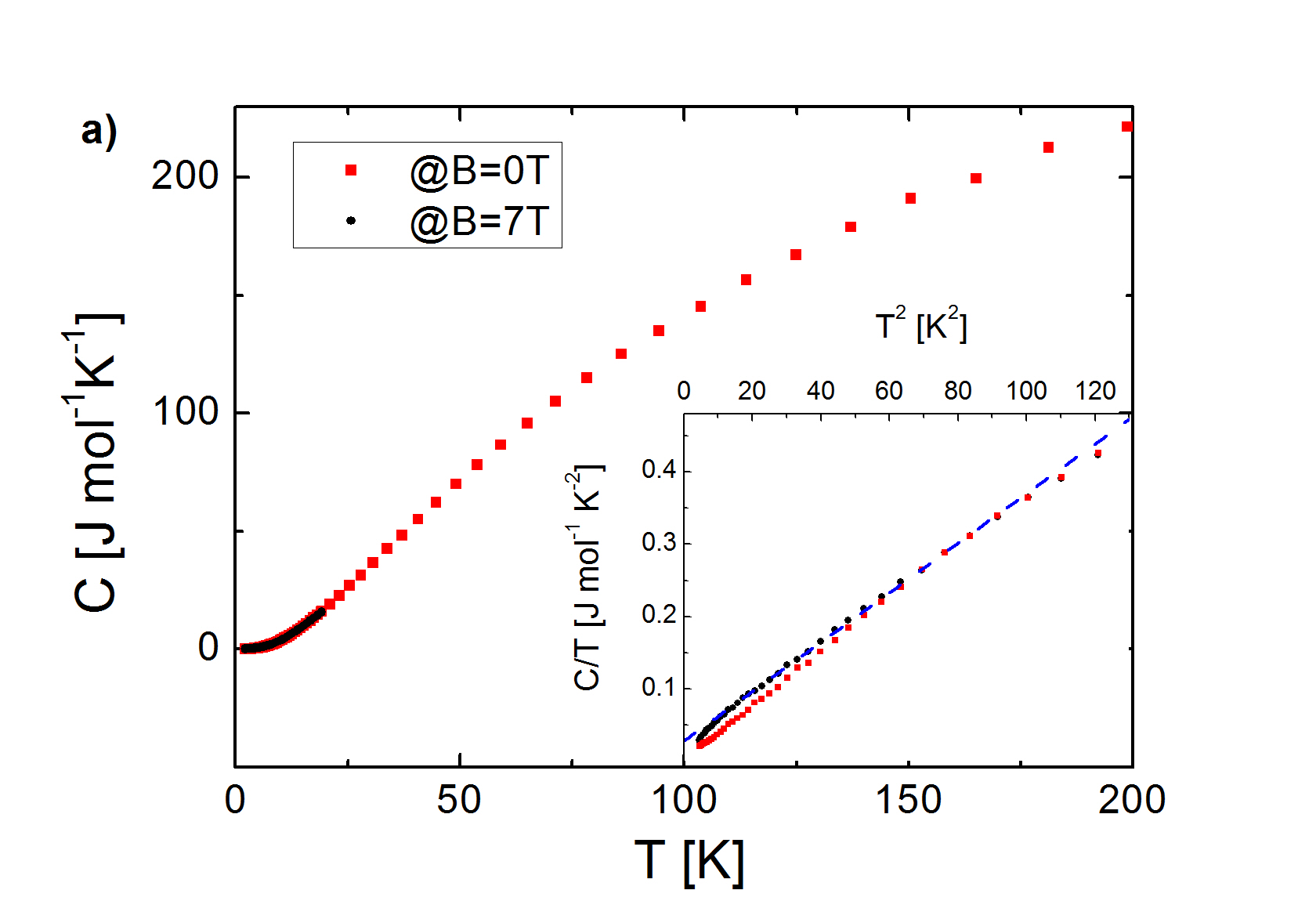}
	\end{minipage}%
	\begin{minipage}{.5\textwidth}
		\centering
		\includegraphics[width=1\linewidth]{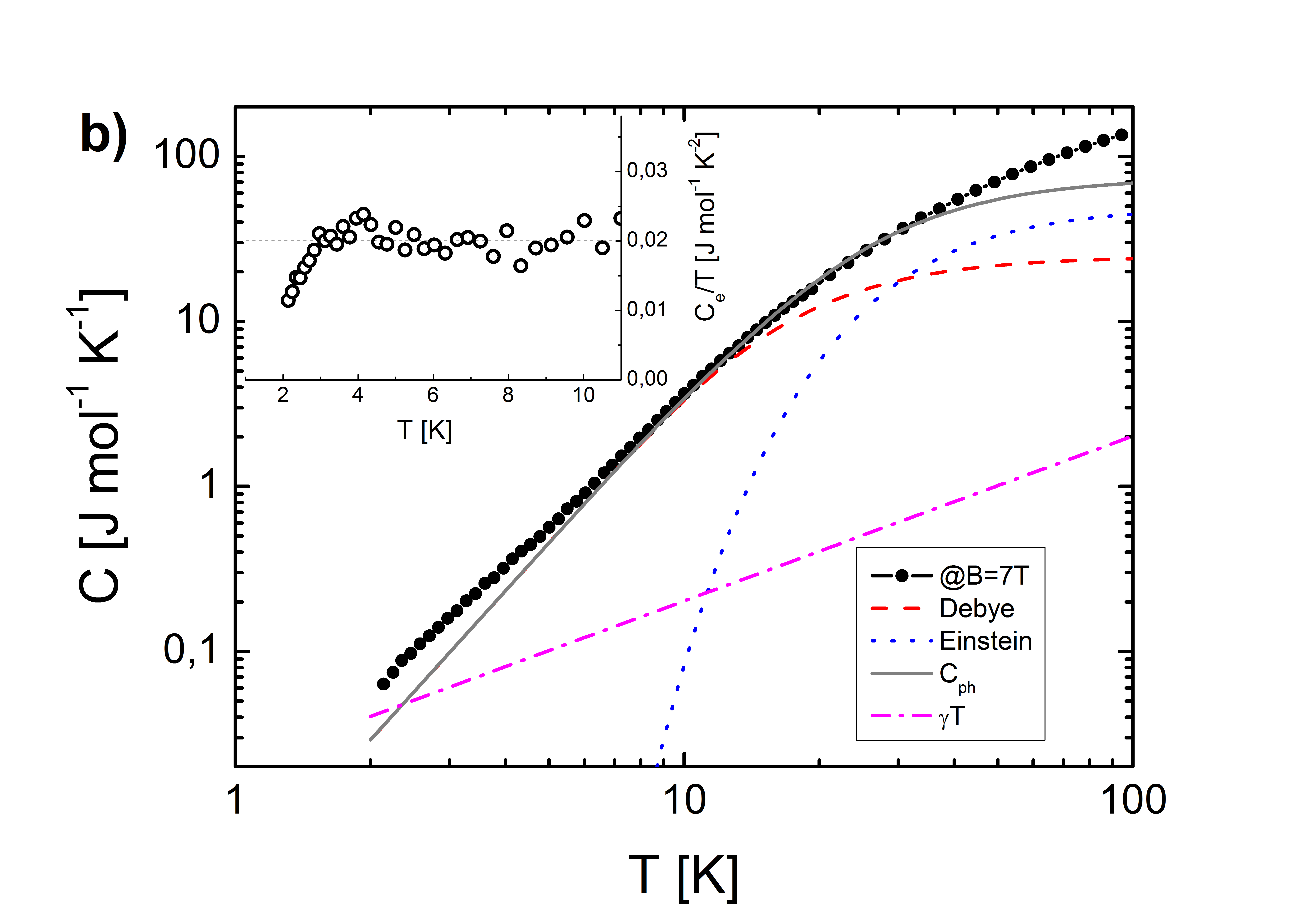}
	\end{minipage}
	\caption{{\bf a$)$} Temperature dependence of $C(0$ T$)$ between 2 and 200 K (red squares) and $C(7$ T$)$ up to 20 K (black dots). Inset: $C(0$ T$)/T$ and $C(7$ T$)/T$ versus $T^2$ from 2 to 11 K. The blue dashed line underlines the linear behaviour of $C(7$ T$)/T$ vs $T^2$ between 3 and 10 K; {\bf b$)$} bi-logarithmic plot of $C(7$ T$)$ (black dots), $C_D$ (red dashed line), $C_E$ (blue dotted line), $C_{ph}$ (grey line) and $C_e$ (magenta dashed-dotted line) between 2 and 100 K. Inset: $C_e/T$ from 2 to 11 K.}
	\label{figA2}
\end{figure}

Figure \ref{figA2}a shows the specific heat $C(0$ T$)$ measured at $B=0$~T in the temperature range 2-200 K (red squares) and at $B=7$~T ($C(7$ T$)$) between 2 and 20 K (black dots). In the inset of figure \ref{figA2}a the ratios $C(0$ T$)/T$ and $C(7$ T$)/T$ versus $T^2$ are reported in the low temperature limit, i.e. from 2 to 11 K. We do not observe any anomaly in $C$ in correspondence of $T_c$. Independently of the quality of the sample (e.g.~Meissner volume), the superconducting anomaly is hardly observable in Bi-based compounds \cite{Collocott1991,junod1999}. However our measurements clearly show that $C(7$ T$)$ departs from $C(0$ T$)$ below 8~K, indicating that the superconducting state is suppressed by the applied magnetic field. 

Figure \ref{figA2}a shows a linear dependence of $C(7$ T$)/T$ versus $T^2$ between 3 and 10 K, whereas the curve rapidly bends at higher temperature. Below 3 K, the deviation from linearity is due to the onset of superconductivity, consistent with resistance measurements in figure \ref{figA1}b where it can be observed that at $B=7$ T, $T_c$ is shifted below 4 K.

In the following we first estimate the phononic contribution $C_{ph}$ by fitting data between 3 and 40 K. We consider the conventional expression $C_{ph}=C_D+C_E$ where $C_D$ and $C_E$ are the Debye and Einstein contributions respectively:
\begin{equation}
 C_D=9R(T/\theta)^3\int_{0}^{\theta/T}\frac{x^4e^x}{(e^x-1)^2}dx, \qquad C_E=3r_ER(T_E/T)^2\frac{e^{T_E/T}}{(e^{T_E/T}-1)^2} \ ,
\end{equation} 
being $R$ the gas constant, $\theta$ the Debye temperature, $T_E$ the Einstein temperature and $r_E$ the degeneracy of the Einstein mode that we set equal to 2 due to the presence of the double BiO layers in the unit cell, as shown in \cite{tajima1991}.

Figure \ref{figA2}b shows the bi-logarithmic plot of the experimental data (black dots) together with the best fit curves $C_D$ (red dashed line), $C_E$ (blue dotted line) and $C_{ph}$ (grey line). It can be observed that $C_{ph}$ reproduces the experimental data between 11 and 35 K. Above 35 K, further optic modes with higher energy \cite{tajima1991} start to contribute. The best fit parameters are $\theta=81$ K and  $T_E=112$ K. The low $\theta$ and the presence of a low-lying optic mode (in accordance with \cite{Collocott1991} and \cite{tajima1991}) account for the bending of $C(7$ T$)/T$ vs $T^2$ above 10 K (figure \ref{figA2}b). The Debye temperature $\theta=$81 K is somehow lower than the value of 228 K reported in ref. \cite{Collocott1991}, but reasonably compatible with that, taking into account that the authors multiplied $\theta$ by a factor $\sqrt[3]{N_{atm}}=$2.2, being $N_{atm}=11$ the number of atoms per unit cell of Bi$_2$Sr$_2$CuO$_6$. 

The normal state electronic contribution was estimated as $C_e=C(7$ T$)-C_{ph}$. Remarkably, $C_e$ is $T$-linear ($C_e=\gamma T$) and we evaluate $\gamma$ value of 20 mJ/(mol K$^2$) (dotted line) from fitting data down to 3 K, where the onset of $T_c$ appears. For comparison, the obtained value is a factor 2 higher than the value reported in ref. \cite{Collocott1991} but we believe that there the $\gamma$ value was underestimated, since no magnetic field was applied to suppress superconductivity. The $\gamma$ value of 20 mJ/(mol K$^2$) is consistent with Sommerfeld coefficients estimated in other metallic phases of cuprate superconductors \cite{plakida2010}. In addition, it is also compatible with the peak values of $\gamma$ which have been recently reported for YBCO, Nd-LSCO and Eu-LSCO.

In figure \ref{figA2}b we plot $C_e$ between 2 and 100 K (magenta dashed-dotted line) and it can be noticed that $C_e\sim$4\% of $C$ at 11 K, becoming negligibly small above this temperature. In conclusion, we point out that the observed $C_e=\gamma T$ directly links to the $T$-linear dependence of the density of entropy $s$ (one of our results, see equation 4 of the main text) by thermodynamic relations, independently of the microscopic description. Remarkably, an entropy linear in temperature is in accordance with the early measurements of the specific heat on optimally doped YBCO performed by Loram et al.~\cite{Loram}.

\end{document}